\documentclass[
pra,
 reprint,
superscriptaddress,
preprintnumbers,
 amsmath,amssymb,
 aps,longbibliography,
]{revtex4-1}

\usepackage{graphicx}
\usepackage{hyperref}
\usepackage{acro}

\usepackage{multirow}
\usepackage[version=4]{mhchem}
\usepackage{dsfont}
\usepackage{float}
\usepackage{caption}
\usepackage{subcaption}
\usepackage{braket}
\usepackage{color}
\usepackage{xcolor}

\hypersetup{
	colorlinks   = true,
	citecolor    = blue,
	linkcolor = blue,
    urlcolor = blue
}
\DeclareAcronym{qmc}{
  short = QMC,
  long  = quantum Monte Carlo,
}
\DeclareAcronym{dmc}{
  short = DMC,
  long  = diffusion Monte Carlo,
}
\DeclareAcronym{vmc}{
  short = VMC,
  long  = variational Monte Carlo,
}

\DeclareAcronym{dft}{
  short = DFT,
  long  = density functional theory,
}
\DeclareAcronym{tddft}{
  short = TDDFT,
  long  = time-dependent density functional theory,
}
\DeclareAcronym{rpa}{
  short = RPA,
  long  = random-phase approximation,
}
\DeclareAcronym{paw}{
  short = PAW,
  long  = projector augmented-wave,
}
\DeclareAcronym{bz}{
  short = BZ,
  long  = Brillouin zone,
}
\DeclareAcronym{ccecp}{
  short = ccECP,
  long  = correlation-consistent effective core potential,
  long-plural = correlation-consistent effective core potentials,
}

\DeclareAcronym{pbe}{
  short = PBE,
  long  = Perdew--Burke--Ernzerhof,
}
\DeclareAcronym{lda}{
  short = LDA,
  long  = local density approximation,
}

\DeclareAcronym{vdw}{
  short = vdW,
  long  = van der Waals,
}
\DeclareAcronym{twod}{
  short = 2D,
  long  = two-dimensional,
}
\DeclareAcronym{hbn}{
  short = h-BN,
  long  = hexagonal boron nitride,
}
\DeclareAcronym{tmd}{
  short = {TMD},
  long  = transition metal dichalcogenide,
}

\DeclareAcronym{icsd}{
  short = ICSD,
  long  = Inorganic Crystal Structure Database,
}
\DeclareAcronym{incite}{
  short = INCITE,
  long  = Innovative and Novel Computational Impact on Theory and Experiment,
}
\DeclareAcronym{alcf}{
  short = ALCF,
  long  = Argonne Leadership Computing Facility,
}
\DeclareAcronym{olcf}{
  short = OLCF,
  long  = Oak Ridge Leadership Computing Facility,
}
\DeclareAcronym{doe}{
  short = DOE,
  long  = U.S. Department of Energy,
}

\DeclareAcronym{supinfo}{
  short = SI,
  long  = Supporting Information,
}

\DeclareAcronym{mad}{
  short = MAD,
  long  = mean absolute deviation,
}
\DeclareAcronym{md}{
  short = MD,
  long  = mean deviation,
}

\begin{document}

% PRX needs a strong title - see recent articles
% Original title:
%\title{Towards a 2D Bilayered Materials Database using Diffusion Monte Carlo}
% Suggestions:
%Too long?
% \title{Binding Energies and Charge Redistribution of Two-Dimensional Van der Waals Bilayers: Benchmark Ab Initio Quantum Monte Carlo Calculations}
\title{Diffusion Quantum Monte Carlo Benchmark of Interlayer Binding and Charge Redistribution in Chemically Distinct Two-Dimensional Van Der Waals Bilayers}
% Other ideas:
%\title{Binding Energies and Charge Redistribution of Two-Dimensional Van der Waals Bilayers: Benchmark Ab Initio Quantum Monte Carlo Calculations}

%\title{Binding Energies and Charge Redistribution of Eleven Two-Dimensional van der Waals Bilayers: Benchmark ab initio Quantum Monte Carlo Calculations and Analysis of Density Functional Approximations}
%Benchmark Quantum Monte Carlo Binding Energies and Charge Redistribution in Two-Dimensional van der Waals Bilayers
%A Many-Body Benchmark for Two-Dimensional van der Waals Interfaces

\author{Kayahan Saritas}
\email{saritask@ornl.gov}
\affiliation{Materials Science and Technology Division, Oak Ridge National Laboratory, Oak Ridge, TN, USA}
\author{Hyeondeok Shin}
\affiliation{Computational Sciences Division, Argonne National Laboratory, Lemont, IL, USA}
\author{Jaron T. Krogel}
\affiliation{Materials Science and Technology Division, Oak Ridge National Laboratory, Oak Ridge, TN, USA}
\author{Anouar Benali}
\affiliation{Computational Sciences Division, Argonne National Laboratory, Lemont, IL, USA}
\author{P. Ganesh}
\affiliation{Center for Nanophase Materials Sciences, Oak Ridge National Laboratory, Oak Ridge, TN, USA}
\author{Paul R. C. Kent}
\affiliation{Computational Sciences and Engineering Division, Oak Ridge National Laboratory, Oak Ridge, TN, USA}
\thanks{This manuscript has been authored by UT-Battelle, LLC under Contract No.\ DE-AC05-00OR22725 with the U.S.\ Department of Energy. The United States Government retains and the publisher, by accepting the article for publication, acknowledges that the United States Government retains a non-exclusive, paid-up, irrevocable, worldwide license to publish or reproduce the published form of this manuscript, or allow others to do so, for United States Government purposes. The Department of Energy will provide public access to these results of federally sponsored research in accordance with the DOE Public Access Plan (\url{https://www.energy.gov/doe-public-access-plan}).}

\begin{abstract}
Interlayer interactions in two-dimensional materials can generate emergent phenomena absent in their constituent monolayers, including unconventional magnetic order, multiferroicity, and topological magnetic phases. Predicting such emergent behavior requires simultaneously resolving long-range dispersion, short-range orbital hybridization, and electronic correlation—interactions that are intrinsically nonlocal and many-body and remain challenging even for advanced density-functional approximations. Here we establish a systematically controlled many-body benchmark for diverse bilayer materials using diffusion Monte Carlo (DMC), spanning single-sheet materials, transition-metal dichalcogenides, and magnetic transition-metal halides. We obtain equilibrium separations, binding energetics, interlayer vibrational modes, and charge redistribution, finding excellent agreement with available experiments while revealing systematic and property-dependent failures across widely used semilocal, meta-GGA, and dispersion-corrected density functionals. Beyond energetics, DMC resolves subtle interlayer charge rearrangements, particularly leading to long-ranged dipolar tails in magnetic Cr trihalides that survive well beyond the regime where semilocal DFT predicts appreciable interlayer polarization, providing a many-body basis for understanding interlayer-coupled ferroic and magnetic phenomena. The resulting energies, response properties, and high-accuracy electron densities constitute a transferable benchmark for developing next-generation density functionals. Finally, we provide a scalable high-performance-computing workflow that enables systematic expansion of many-body benchmark datasets across emerging families of layered quantum materials by the scientific community.
\end{abstract}

\maketitle

\section{Introduction}
The emergence of 2D materials has revolutionized condensed matter physics and materials science, offering unprecedented opportunities for both fundamental research and technological applications \cite{Novoselov2005,Geim2013,Manzeli2017}. These materials, which include graphene, \ac{hbn}, and \acp{tmd}, exhibit a range of novel electronic, optical, and mechanical properties that make them suitable for applications in nanoelectronics, optoelectronics, and quantum information science
\cite{Fiori2014,Mak2016,Liu2019}. Beyond monolayers, layered structures introduce an additional degree of freedom in material design through interlayer interactions, stacking arrangements, and electronic band hybridization, leading to tunable properties such as superconductivity, ferroelectricity, and topological states \cite{Novoselov2016,Duong2017}.
Interlayer sliding and stacking can break inversion symmetry and produce switchable out-of-plane polarization, as in sliding ferroelectric bilayers, while magnetic stacking and crystal symmetry likewise control exchange pathways in 2D magnets and in related phenomena such as altermagnetism.
Accurate interlayer separations, binding curves, and charge redistribution are therefore prerequisites for modeling these stacking-tunable polar and magnetic responses.
% \pkcom{Need to add modern citations on twistronics, moire etc. and recent papers from PRX and the 2020s}

The accurate theoretical characterization of layered materials remains a challenging task because several distinct physical ingredients coexist. 
Non-local correlation is the electron correlation that depends on spatially separated regions of the density, hence missing in the local and semilocal exchange--correlation functionals. 
\ac{vdw} interactions are the long-range, asymptotically decaying part of that non-local correlation and provides the majority of the attraction between weakly interacting layers. 
These interactions are introduced to standard \ac{dft} functionals via the addition of pairwise interactions\cite{Grimme2010,Tkatchenko2009} up to self-consistent non-local \ac{vdw} functionals with minimal empirical input\cite{Dion2004,Klimes2011,Ambrosetti2014,Sabatini2013}. 
More sophisticated, beyond-DFT, approaches based on \ac{rpa} provide an explicit treatment of long-range correlation effects \cite{Harl2010}; however, \ac{rpa} atomization energies for solids are known to be systematically overestimated by up to 15\% \cite{Harl2010}, and its quantitative reliability for layered and weakly bonded 2D systems remains to be validated using other many-body methods.
An additional challenge is interlayer charge transfer at short distances in certain bilayers.
As a consequence, the accuracy of widely used methods for interlayer interactions in 2D bilayers is not well-established, and the absence of systematically controlled reference data hinders both methodological assessment and further theoretical development. 

This has motivated growing interest in \ac{qmc} applications, in particular \ac{dmc}, which enables highly accurate total energy calculations in molecular and periodic systems through an explicit treatment of electron--electron correlations \cite{FOU2001} with few approximations. By describing intra- and interlayer interactions on an equal footing without empirical correction, \ac{dmc} provides a natural framework for establishing a many-body benchmark for interlayer binding in layered 2D materials\cite{WinesReview2025}. Improvements in the methods, their implementation, and computational resources, now enables the relatively small binding energies to be well resolved.

To move beyond isolated case studies and expose systematic trends, we selected 10 bilayer systems that span the main families of layered 2D materials while remaining accessible to a consistent high-throughput workflow based on bulk crystal structures from \ac{icsd} \cite{Bergerhoff1987}. Bilayer graphene and \acl{hbn} provide canonical, weakly bound sp$^2$-bonded benchmarks with extensive prior many-body and experimental data, including prior \ac{dmc} studies of bilayer graphene \cite{Mostaani2015,Spanu2009}. Six transition-metal dichalcogenides, MoX$_2$ and WX$_2$ with X = S, Se, and Te, cover technologically relevant semiconductors and introduce systematic variation in atomic mass, polarizability, and interlayer binding strength. Chromium trihalides, CrI$_3$ and CrBr$_3$, in both antiferromagnetic and ferromagnetic interlayer alignments, extend the benchmark to magnetically ordered, correlated insulators for which mean-field \ac{dft} plus empirical dispersion corrections are expected to be the least reliable. 

Together, these systems probe a wide range of interlayer separations and binding energies providing a stringent and chemically diverse test bed for assessing whether \ac{dft}-based approaches can reproduce many-body reference interlayer energetics.Fixed-node \acs{dmc} has become a standard reference for weak and strong interlayer coupling in \acs{twod} systems when \acs{dft} dispersion treatments disagree \cite{WinesReview2025}.
While bilayer graphene \cite{Mostaani2015, Spanu2009}, bilayer hexagonal boron-nitride \cite{Hsing_2014} and their heterostructures \cite{szyniszewski2025adhesion} and bilayer MoS$_2$ \cite{huang2026bilayers} were previously studied using \acs{dmc}, the study of layered materials with \acs{dmc} has been often limited to monolayers, layered bulk materials and their defects \cite{Wines2023, wines2025quantum, Ichibha2021, Staros2026, saritas2025increased, ghaffar2026critical, wines2021gasxse, huang2024mos2, ahn2026inse, ahn2025ptse2, Shulenburger2015, frank2019phosphorene, ahn2021metallic, hunt2020hbn,thomas2022defects, GaneshJCTC2014, ahn2020borophene, ahn2026arsenene, huang2023phosphorene}. 
% \kscom{added a few references}
% \pkcom{Need to add some of Ganesh JCTC, Shulenberger, Drummond, Kwon, Stich, other recent QMC papers from the 2020s (likely reviewers!). See Wines 2025 Review.}
% \pkcom{We can also cite more of our recent work here, but not too much!}

In this work, we use fixed-node \ac{dmc} to systematically benchmark interlayer physics in chemically distinct 2D bilayers. We benchmarked the binding energy and interlayer separation at its minima ($E_{\mathrm{b}}$, $d_0$), out-of-plane breathing mode frequencies $\omega$ and the layer-resolved dipole moments extracted from plane-averaged charge densities $\rho(z)$. 
We aim to establish a reliable benchmark dataset that can aid in the validation of approximate computational methods. This database will not only serve as a reference for researchers exploring new bilayered materials but also provide insights into the fundamental interactions governing their stability and functionality.
Tabulated energetics and the raw charge densities are available in the supplementary information \cite{supp} and in associated data repositories. 
\begin{figure*}
  \centering
  \includegraphics[width=\linewidth]{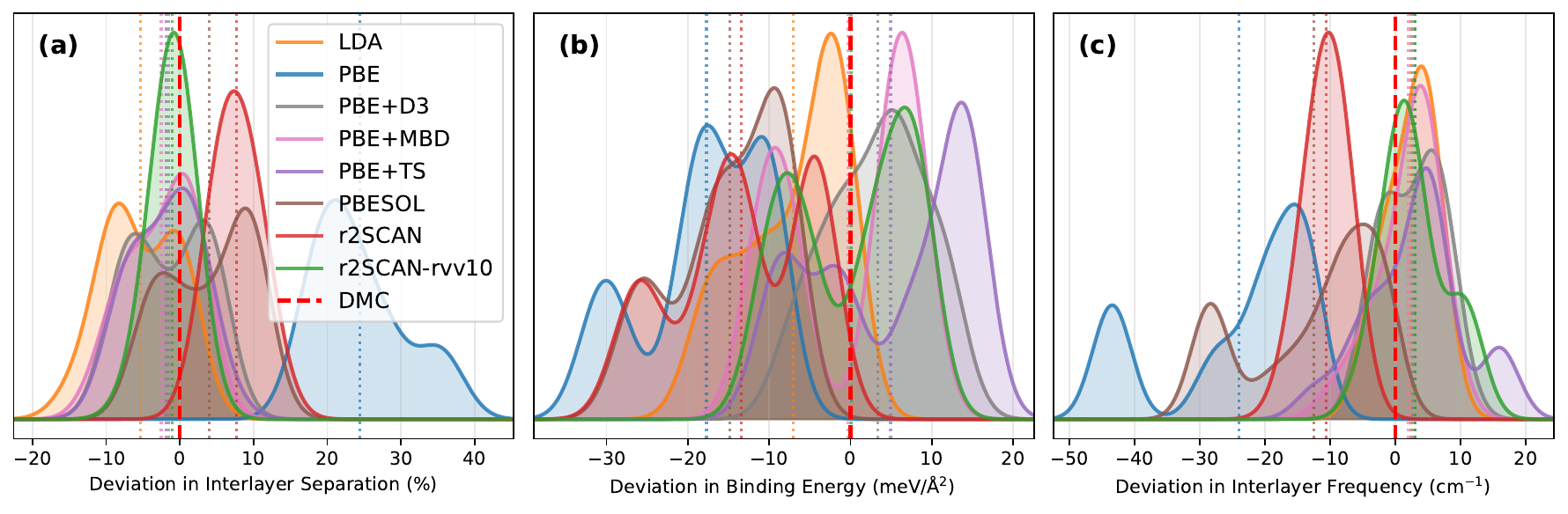}
  \caption{Deviations of \ac{dft} functionals from \ac{dmc}, in aggregate across all materials, for (a) equilibrium interlayer separation $d_0$ (percent), (b) interlayer binding energy $E_{\mathrm{b}}$ per unit area (meV/\AA$^2$), and (c) out-of-plane breathing-mode frequency $\omega$ (from the curvature of $E_{\mathrm{b}}(d)$ at $d_0$). Absolute $d_0$ and per-atom $E_{\mathrm{b}}$ values are given in \acs{supinfo} Tables~S1--S2.}
  \label{fig:dft_vs_dmc}
\end{figure*}

\section{Methods}
\subsection{Utilized geometries}
We use a top-down approach to generate layered structures from bulk materials in the \ac{icsd} \cite{Bergerhoff1987}. Therefore, the stacking order in the bilayer is transferred from the stacking order in the bulk. Detailed information on the structures used and their stackings are provided in the \ac{supinfo}\cite{supp}. The monolayer and bilayer geometries used for the interlayer binding curves are consistently obtained from the bulk experimental coordinates. For direct comparison, the same geometries were used for the \ac{dft} binding curves: in-plane lattice constants and intralayer atomic coordinates are taken from the bulk-derived structures and held fixed while the interlayer separation $d$ is scanned rigidly. For \ac{dft} calculations, interlayer separations were studied up to 8~\AA, while in \ac{dmc} they were studied up to 5~\AA. The use of periodic boundary conditions in z-direction means that image interactions are possible. We used a vacuum layer of at least 12~\AA\ to minimize the image interactions. 

\subsection{Calculated quantities}
We defined the interlayer separation ($d$) of the bilayer as the minimum vertical distance between the atoms in each layer of the bilayer. Using this definition, we fitted
\begin{equation}
E_{\mathrm{b}}(d)=A_2d^{-2}+A_4d^{-4}+A_6d^{-6}+A_8d^{-8}
\label{eq:binding}
\end{equation}
to the binding energies calculated in \ac{dft} and \ac{dmc}. We have extensively tested different functional forms including exponentials and other polynomials (\acs{supinfo} Fig.~S2 and Table~S4). We found that using the 6--12 Lennard-Jones potential produced reasonable fits ($R^2>0.9$) for more repulsive \ac{dft} functionals (PBE, r$^2$SCAN and PBEsol with no dispersion correction), but failed ($R^2<0.7$) for attractive \ac{dft} functionals (LDA, \acs{dft}+\acs{vdw}) and \ac{dmc}. 
Additional terms were added to improve the quality of the fit.
We find that using these four parameters yields observable quantities that are robust with respect to the choice of fitting function.
The form above was used to extract all interlayer separations $d_0$, binding energies $E_{\mathrm{b}}$, and out-of-plane breathing-mode frequencies, $\omega$,
reported in this work. 
This functional form is related to earlier \ac{dmc} fits for bilayer graphene \cite{Mostaani2015}, where $E_{\mathrm{b}}'(d)=A_2d^{-4}+A_4d^{-8}+A_6d^{-12}+A_8d^{-16}$ was found to give reliable interlayer separations and phonon breathing-mode frequencies. 
In our tests, the Eq.~\ref{eq:binding} also produces high quality fits ($R^2>0.99$ for most materials and functionals; mean $R^2\approx0.99$ for \ac{dmc}). 
Previous \ac{dmc} and \ac{rpa} calculations on graphene predicted that for the long-range portion of the binding-energy curve the dominant terms have $d^{-\alpha}$ scaling with $\alpha < 4$, which agrees with our choice of fitting function \cite{Spanu2009, Gould2008}.

We calculated the binding energy as $E_{\mathrm{b}}=E_{\mathrm{bilayer}}-2\times E_{\mathrm{monolayer}}$, where $E_{\mathrm{bilayer}}$ and $E_{\mathrm{monolayer}}$ are the total energies of the bilayer and the corresponding isolated monolayer, respectively.
Tabulated values in the \acs{supinfo} are reported per atom (meV/atom) \cite{supp}.
For cross-material comparison in Fig.~\ref{fig:dft_vs_dmc}(b) we further normalize $E_{\mathrm{b}}$ by the in-plane cell area (meV/\AA$^2$).
To obtain phonon breathing modes, the second derivative of the fitted binding-energy curve, $k$, is evaluated at the equilibrium separation $d_0$.
The reduced mass of the bilayer reduces to $\mu=\braket{m}/2$, where $\braket{m}$ is the average mass per atom, which can be written as $\braket{m}=(nw_A+mw_B)/(n+m)$ for A$_n$B$_m$. The breathing-mode frequency (spectroscopic wavenumber) is calculated as
\begin{equation}
\omega=\frac{1}{2\pi c}\sqrt{\frac{k}{\mu}},
\end{equation}
where $c$ is the speed of light, so that $\omega$ is reported in cm$^{-1}$ for comparison with Raman experiments.
In \ac{dmc}, the mean and the uncertainty of all quantities are obtained through Monte Carlo bootstrapping. 
The equilibrium $d_0$, $E_{\mathrm{b}}$, and $\omega$ summarized in Fig.~\ref{fig:dft_vs_dmc} are extracted from these fitted curves; representative examples are shown in Fig.~\ref{fig:binding_curves}, with the full benchmark set in \acs{supinfo} Fig.~S1 \cite{supp}.

\subsection{Diffusion Monte Carlo calculations}
Fixed-node many-body \ac{dmc} calculations \cite{FOU2001} were performed as implemented in \texttt{QMCPACK} v4.0.0 \cite{KimJPCM2018,KentJCP2020} via the \texttt{Nexus} workflow automation package \cite{KRO2016}. 
For the \ac{qmc} calculations, single-determinant Slater-Jastrow type trial wavefunctions \cite{SLA1929, JAS1955} were generated using \ac{ccecp} potentials \cite{ANN2018, BEN2017, WAN2019, WAN2022} via the \texttt{Quantum ESPRESSO} package v7.4.0 \cite{GIA2009} using the \ac{pbe} \cite{Perdew1996} exchange-correlation functional for all materials.
We used T-moves approximation for variational estimation of non-local pseudopotential terms \cite{Casula2006}.
For magnetic Cr-halides, the \ac{dmc} ground state was obtained from PBE$+U$ trial wavefunctions \cite{ANI1991} with the Hubbard correction applied to Cr~$3d$ states. We used a Hubbard $U$ of 1.0 eV for CrI$_3$ and CrBr$_3$ which minimizes the fixed-node DMC energy in their monolayers \cite{supp}. 
% \kscom{add to SI}
The single particle orbitals were represented in a hybrid basis set~\cite{LuoJCP2018} to reduce memory consumption.
We used two and three body Jastrow terms \cite{Drummond2004} to optimize the trial wavefunction. These Jastrow parameters were optimized via mixed energy/reweighted variance minimization \cite{UMR2007}.
A \ac{dmc} time step of 0.005~Ha$^{-1}$ was used to minimize imaginary-time-step errors, and \ac{dmc} total energies were extrapolated to the infinite-size limit using linear extrapolations over $N^{-5/4}$ in the number of atoms $N$, as previously demonstrated to be appropriate for 2D materials \cite{Drummond2008}. 
We used supercells constructed from 2$\times$2 scalar matrices (N$\times$N) where N is an integer in the 2-4 range acting upon the primitive cells of each material to extrapolate the finite size errors. For bilayer graphene and hBN we used supercells containing up to 256 atoms, 1024 electrons, for \acp{tmd} up to 96 atoms, 832 electrons and for Cr-trihalides we used up to 144 atoms and 1260 electrons. 
We used twist averaged boundary conditions to reduce residual one-body finite-size errors in the Brillouin zone. 
Regular twist mesh grids were converged in the primitive cells using DFT calculations, then commensurate twist grids were used for all supercells.
In this dataset the resulting statistical errors are typically ${\sim}0.01$~\AA\ on $d_0$, ${\sim}0.1$--1~meV/atom on $E_{\mathrm{b}}$, and 1-5 cm$^{-1}$ on $\omega$, comparable to or smaller than the \ac{dft}--\ac{dmc} discrepancies discussed below.

\subsection{Density Functional Theory Calculations}
\Ac{dft} calculations for the structural and energetic benchmarks were performed in \texttt{VASP} v5.4.4 \cite{Kresse1996,Kresse1996b} using \ac{paw} potentials \cite{Blochl1994}.
For charge density benchmarks, in order to be consistent with the \ac{dmc} calculations, we used the same \ac{ccecp} potentials \cite{ANN2018, BEN2017, WAN2019, WAN2022} and settings as for the DMC trial wavefunction generation, changing only the exchange--correlation functional and any dispersion correction.
Geometries follow the rigid $d$ scan described above; only the exchange--correlation functional and dispersion correction are varied.
We used LDA\cite{PerdewPRB1981}, PBE\cite{Perdew1996}, PBEsol\cite{Perdew2008} and r$^2$SCAN\cite{Sun2015,Furness2020} functionals and D3\cite{Grimme2010}, TS \cite{Tkatchenko2009}, MBD\cite{Ambrosetti2014}, and rVV10 \cite{Sabatini2013} \ac{vdw} correction methods.
D3, TS and MBD corrections were only paired with PBE, which otherwise lacks long range dispersion, while the rVV10 correction was only paired with r$^2$SCAN.
A kinetic energy cutoff of 520~eV was used, with a reciprocal-grid density of 350~per~\AA$^{-3}$ of the \ac{bz} (a $12\times12\times1$ grid for graphene).

\section{Results and Discussion}
We summarize the results for equilibrium interlayer separation $d_0$, binding energy $E_{\mathrm{b}}$, and out-of-plane breathing-mode frequency $\omega$ in Figs.~\ref{fig:dft_vs_dmc} and~\ref{fig:binding_curves}. All individual results are given in \acs{supinfo} Tables~S1--S3 \cite{supp}. For every quantity below, we also provide a comparison between DMC and high-level ab initio calculations and experimental measurements from the literature. We find that our DMC results are in good agreement with the prior high-level ab initio calculations. Care must be taken when comparing any of the theoretical results to experimental data due to, e.g., the use of fixed lattice constants and coordinates.

For each observable $X \in \{d_0, E_{\mathrm{b}}, \omega\}$, we define per-bilayer deviations
$\Delta X_i = X_i^{\mathrm{DFT}} - X_i^{\mathrm{DMC}}$ over ten bilayers ($N=10$).
The signed \ac{md} is $\mathrm{MD} = \frac{1}{N}\sum_i \Delta X_i$;
the \ac{mad} is $\mathrm{MAD} = \frac{1}{N}\sum_i |\Delta X_i|$.
Values quoted in Results are from \acs{supinfo} Tables~S1--S3 unless noted \cite{supp}. 

\subsection{Equilibrium interlayer separation}
Semilocal functionals exhibit a systematic ordering of equilibrium interlayer separation $d_0$: LDA systematically underestimates by \acs{md}/\acs{mad} of $-0.26/0.26$~\AA, PBE systematically overestimates by \acs{md}/\acs{mad} of $0.86/0.86$~\AA, and PBEsol lies between LDA and PBE, tracking \ac{dmc} more closely with \acs{md}/\acs{mad} of $0.10/0.16$~\AA.
%\pkcom{MD and MAD don't appear to be defined despite the macros?}
We also find that PBEsol performs better for multi-atom-thick layers (e.g., \acp{tmd}) than for single-sheet graphene and \acs{hbn}, consistent with its design for slowly varying densities in solids and slabs \cite{Perdew2008}.
The standard deviation of the signed errors $\Delta d_0$ remains similar for all three semilocal functionals (${\sim}0.14$--$0.17$~\AA; Fig.~\ref{fig:dft_vs_dmc}(a)).
This can indicate that for PBEsol, while the systematic bias is cancelled out, the spread of the errors is still significant and does not indicate a significant improvement over LDA and PBE. 
While r$^2$SCAN also yields improved $d_0$ values, compared to LDA and PBE, it performs worse than PBEsol in terms of the \acs{md} error. However, the standard deviation of the signed errors in r$^2$SCAN is smaller (${\sim}0.06$~\AA) than that of PBEsol (${\sim}0.16$~\AA), as indicated with the sharper peak in Fig.~\ref{fig:dft_vs_dmc}(a). 

The average results in Fig.~\ref{fig:dft_vs_dmc} show that dispersion-corrected functionals markedly improve $d_0$ for all the DFT functionals they are paired with.
Most importantly, r$^2$SCAN--rVV10 achieves an impressive agreement with \ac{dmc} with \acs{mad} of 0.05~\AA, while PBE-based dispersion functionals perform qualitatively similarly (\acs{mad} ${\sim}0.15$~\AA).
Dispersion corrections based on PBE significantly improve the mean and mean-absolute deviation error metrics; however, the standard deviation of the signed errors in all dispersion-corrected PBE functionals is only minimally improved compared to bare PBE, and remains larger than that of bare r$^2$SCAN.
In comparison, the addition of the rVV10 correction to r$^2$SCAN significantly improves every error metric we investigated compared to bare r$^2$SCAN.

Our DMC results for $d_0$ are in good agreement with high-level \textit{ab initio} calculations and experimental measurements.
For bilayer graphene, we find $d_0=3.38(1)$~\AA, which compares well with previous \ac{dmc} of bilayer graphene [3.43~\AA \cite{Mostaani2015}] and bulk graphene [3.36~\AA \cite{Spanu2009}].
\Ac{rpa} predicts similar $d_0$ values for bilayer graphene (3.39~\AA \cite{Olsen2013}) and bulk graphene (3.34~\AA \cite{Lebegue2010}), close to the experimental bulk spacing of 3.36~\AA \cite{Trucano1975}.
Similar agreement is found for bilayer \acs{hbn} with $d_0=3.36(1)$~\AA, consistent with previous \ac{dmc} (3.25--3.50~\AA \cite{Hsing_2014}) and M{\o}ller--Plesset perturbation theory (3.34~\AA \cite{Constantinescu2013}) and with the experimental bulk spacing of 3.35~\AA \cite{Budak20104702}.
For \acp{tmd} and Cr-halides, our \ac{dmc} vdW gaps lie within ${\sim}0.1$~\AA\ of experimental nearest-layer spacings \cite{Zhan2012,Petkov2002,Schutte1987207,Huang2017,Chen2019CrBr3,McGuire2015}.
Direct comparison with \ac{rpa} bilayer TMDs requires a definition caveat: He \textit{et al.}\ report the metal--metal (or $c/2$) distance, 6.27~\AA\ for AA$'$ MoS$_2$ \cite{He2014}, whereas our $d_0$ is the chalcogen--chalcogen vdW gap.
Subtracting the S--S thickness of one MoS$_2$ layer in our ICSD-derived geometry \cite{Petkov2002}($t=3.130$~\AA) converts that \ac{rpa} value to a gap of 3.14~\AA, in agreement with our 3.13(1)~\AA.
Qualitative trends are also recovered: larger terminating atoms give larger $d_0$ in the TMD series, and ferromagnetic Cr-halide stacking has a larger \ac{dmc} $d_0$ than antiferromagnetic stacking.

\subsection{Interlayer binding energy}
\begin{figure}
  \centering
  \includegraphics[width=\linewidth]{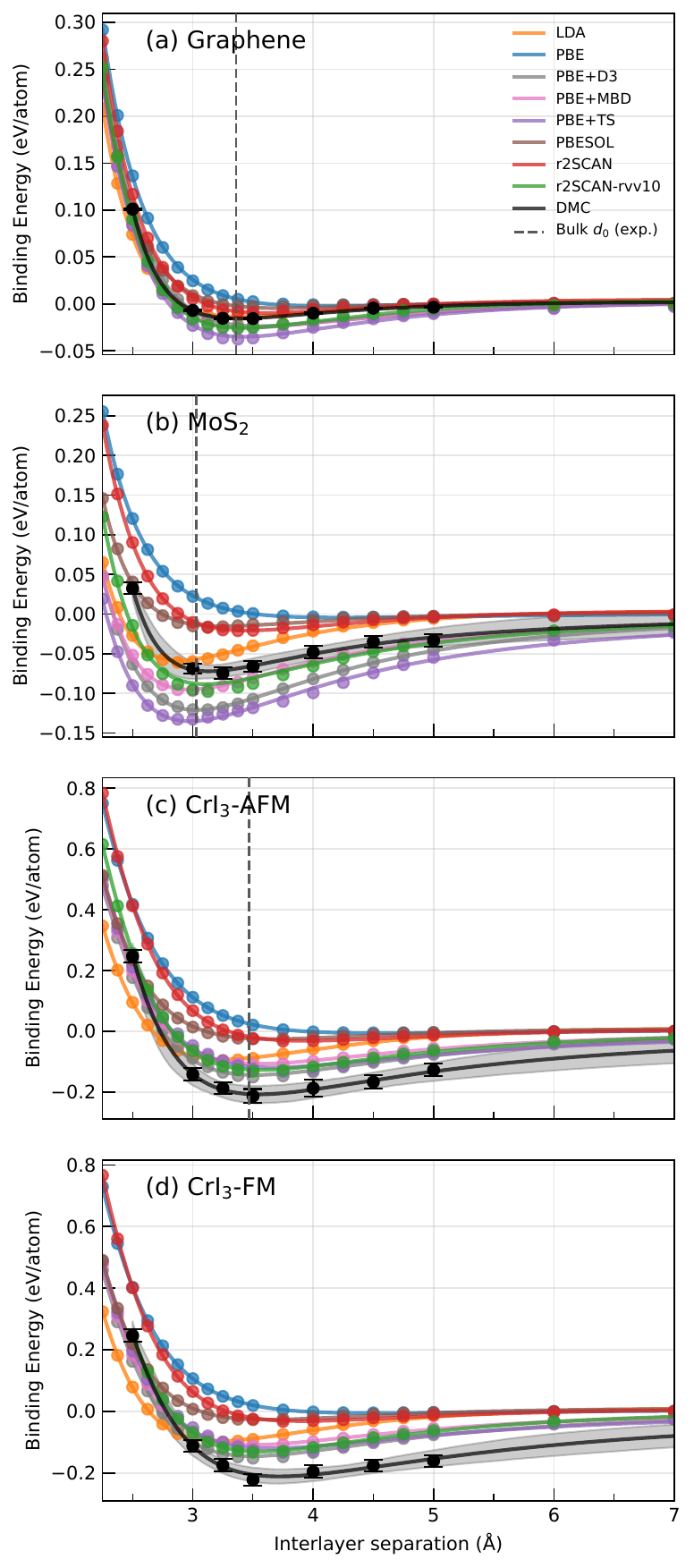}
  \caption{Representative interlayer binding-energy curves $E_{\mathrm{b}}(d)$ and fitted potentials for (a) graphene, (b) MoS$_2$,
  and (c,d) CrI$_3$ in antiferromagnetic and ferromagnetic interlayer alignments. Points are \acs{dft} and \acs{dmc} data; solid
  lines are fits to the form described in Methods. The \acs{dmc} fit is shown with a 95\% confidence band. Curves for all ten
  bilayers appear in \acs{supinfo} Fig.~S1 \cite{supp}. Experimental bulk $d_0$ are reported from ref. \cite{Trucano1975, Petkov2002, McGuire2015}.}
  \label{fig:binding_curves}
\end{figure}

Fig.~\ref{fig:binding_curves} shows representative $E_{\mathrm{b}}(d)$ curves and fits used to extract the benchmark observables in Fig.~\ref{fig:dft_vs_dmc}.
Panel (a) for bilayer graphene shows a shallow semilocal minimum and deep dispersion-corrected wells that overbind relative to \acs{dmc} at the fitted minimum.
Panel (b) for MoS$_2$ illustrates the contrasting curvature of the binding curves: PBE yields a very shallow well, whereas LDA and \acs{dmc} produce deeper minima at shorter separations.
Panels (c) and (d) for antiferromagnetic and ferromagnetic CrI$_3$ show that magnetic stacking order shifts the \acs{dmc} minimum to larger $d$ in the FM case, whereas semilocal and dispersion-corrected \acs{dft} curves fail to reproduce this ordering (curves for all ten bilayers plus two magnetic configurations are given in \acs{supinfo} Fig.~S1 \cite{supp}).

The semilocal ordering of $d_0$ and $E_{\mathrm{b}}$ can be understood from how each functional treats exchange--correlation in regions of weak interlayer density overlap.
LDA, based on the uniform electron gas, underestimates exchange repulsion in inhomogeneous systems and favors interlayer density overlap, yielding the shortest $d_0$.
PBE enhances exchange repulsion through gradient corrections designed for molecular atomization, suppressing interlayer attraction and producing the largest $d_0$ and the weakest binding.
PBEsol partially restores the gradient expansion appropriate for slowly varying solid-state densities, reducing PBE's excessive repulsion while remaining between LDA and PBE on both observables.
%These exchange-driven trends in $d_0$ should not be confused with accurate \ac{vdw} cohesion:
Even where LDA produces short
separations and deeper wells than PBE, the fitted $E_{\mathrm{b}}$ of LDA (and of other semilocal functionals) still typically
falls below \ac{dmc} for \acp{tmd} and Cr trihalides because nonlocal correlation is missing (\acs{supinfo} Table~S2 \cite{supp}).
Graphene and \acs{hbn} are partial exceptions where LDA $E_{\mathrm{b}}$ can approach \ac{dmc} by coincidence of overestimated overlap attraction and underestimated dispersion.
%Tuning the exchange enhancement in semilocal functionals can improve equilibrium geometries, but accurate binding energies require explicit treatment of nonlocal correlation.

Comparison with experimental quantities must be made with care: detailed visual representations of the differences between cleavage, exfoliation and cohesive energy can be found in the
Supplementary Information of Ref. \cite{Bjorkman2012}. While these quantities are usually of similar magnitude, these terms will
differ when the interactions between second and further neighbor planes are significant, as in the case of graphene/graphene
\cite{Gould2013}. Bulk interlayer binding of graphene---the energy to separate the crystal fully into monolayers---was reported as
56(5)~meV/atom in \ac{dmc} \cite{Spanu2009} and 48~meV/atom in \ac{rpa} \cite{Lebegue2010}. Cleavage energy (one basal cut between
two semi-infinite stacks) is typically ${\sim}15$--$20\%$ larger than the binding energy in model estimates
\cite{Girifalco1956,Zacharia2004,Gould2013}: Zacharia et al.\ inferred $61\pm5$~meV/atom from desorption-based exfoliation data
\cite{Zacharia2004}, while Wang et al.\ later measured ${\sim}64$~meV/atom for AB-stacked graphene \cite{Wang2015}. Exfoliation and
related cohesion/adhesion estimates for graphene---$43\pm5$ \cite{Girifalco1956}, $52\pm5$ \cite{Zacharia2004}, and
${\sim}54$~meV/atom from G/G adhesion \cite{Rokni2020}---fall in the same ${\sim}45$--$55$~meV/atom range as the bulk
\ac{dmc}/\ac{rpa} binding energies above, supporting that multilayer cohesion is much stronger than freestanding-bilayer binding.
Our bilayer graphene $E_{\mathrm{b}}$ is 15.4(4)~meV/atom, consistent with bilayer \ac{dmc} [17.7(9)~meV/atom \cite{Mostaani2015}].
For MoS$_2$, \ac{rpa} bilayer binding in the AA$'$ stacking is ${\sim}27$~meV/atom \cite{He2014}, close to our 24.3(1)~meV/atom.
Although we expect the changes to be small, and not yet practical to perform for all the materials, a full relaxation of all the
geometries in \ac{dmc} could further improve the agreement with experiment.

\begin{figure*}[htbp!]
  \centering
  \includegraphics[width=\textwidth]{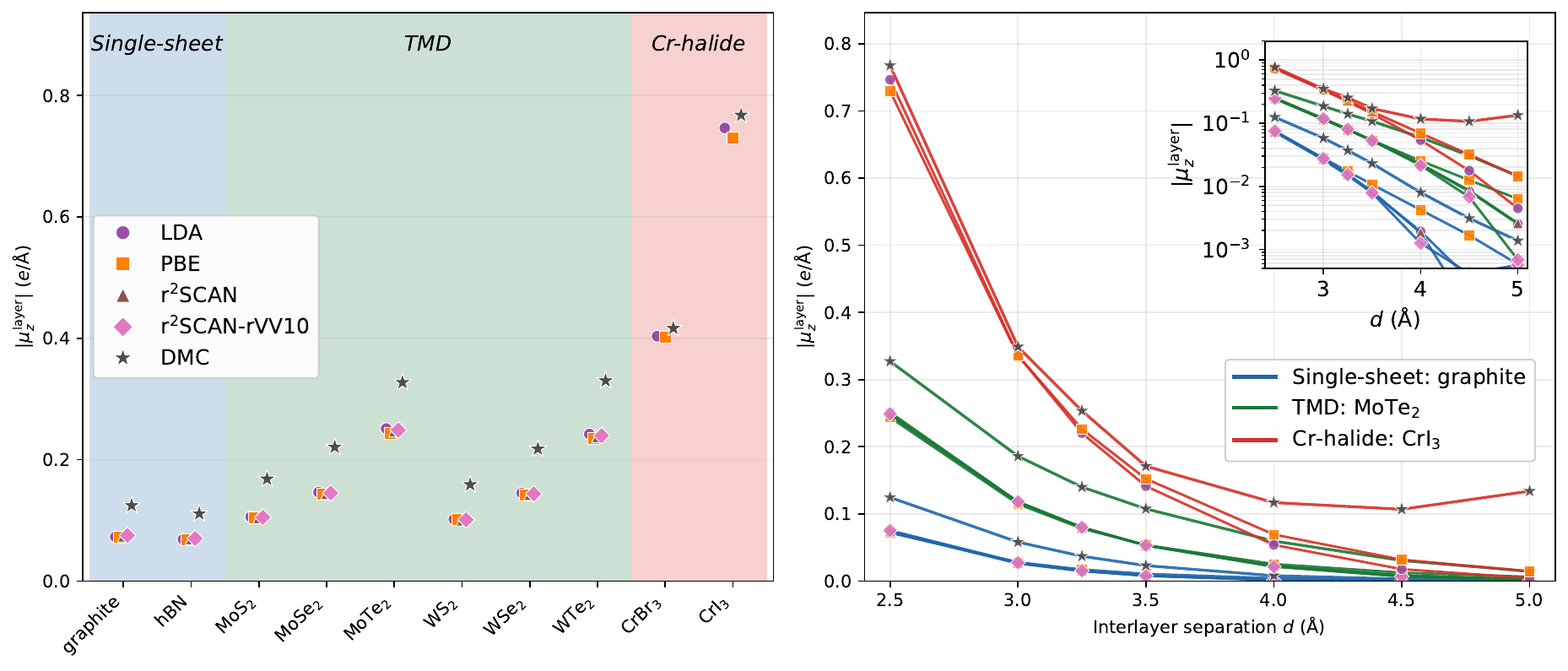}
  \caption{
  Bottom-layer interlayer dipole magnitude $\lvert\mu_z^{\mathrm{layer}}\rvert$ from bilayer $\rho(z)$ (units: $e$\,\AA\ per unit area).
\textbf{(a)} Values at $d = 2.5$~\AA\ for representative \ac{vdw}, \acs{tmd}, and Cr halide bilayers; markers show LDA, PBE, meta-GGA functionals, and \acs{dmc} (see legend).
Shaded bands and labels denote material classes.
\textbf{(b)} Separation dependence for selected bilayers; color denotes material (legend) and line style denotes method (inset legend).
Semilocal \acs{dft} predicts rapid decay for all classes, whereas \acs{dmc} retains a long-ranged tail unique to Cr halides at $d \gtrsim 4$~\AA.}
  \label{fig:dipole}
\end{figure*}

\subsection{Interlayer breathing modes}
Out-of-plane breathing-mode frequencies $\omega$ probe the curvature of $E_{\mathrm{b}}(d)$ at $d_0$ and therefore provide a complementary test to the location and depth of the minimum (Methods; Fig.~\ref{fig:dft_vs_dmc}(c); \acs{supinfo} Table~S3 \cite{supp}).
As for $d_0$, bare semilocal functionals systematically order $\omega$: LDA yields the stiffest wells, PBE the softest, and PBEsol/r$^2$SCAN intermediate values.
PBE underestimates $\omega$ by tens of cm$^{-1}$ across the set (\acs{mad} 24~cm$^{-1}$), consistent with its shallow binding curves [Fig.~\ref{fig:binding_curves}].
LDA, dispersion-corrected PBE, and r$^2$SCAN--rVV10 all yield similar $\omega$ frequencies with \acs{md} less than ${\sim}2.2$ and \acs{mad} less than ${\sim}6$~cm$^{-1}$. In comparison, PBE, PBEsol, and r$^2$SCAN have systematic errors with \acs{md} of ${\sim}{-}24$, ${\sim}{-}13$, and ${\sim}{-}12$~cm$^{-1}$, respectively.

Dispersion corrections restore much of the missing curvature: PBE-D3, PBE-MBD, and r$^2$SCAN--rVV10 reduce the \acs{mad} on $\omega$ to ${\sim}3$--4~cm$^{-1}$, comparable to LDA's fortuitous agreement.
Improvements are not uniform, however---PBE-TS overestimates graphene $\omega$ (89 vs.\ \ac{dmc} 73~cm$^{-1}$), reflecting the same overbinding seen in $E_{\mathrm{b}}$.
In Cr trihalides, large \ac{dft}--\ac{dmc} gaps in well depth and shape propagate into $\omega$: semilocal functionals underestimate AFM CrBr$_3$ frequencies, and magnetic stacking (AFM vs.\ FM) changes \ac{dmc} $\omega$ more than most dispersion-corrected functionals predict.
Overall, $\omega$ is more challenging to predict and functionals that partially correct $d_0$ can still fail for curvature-dependent observables as chemistry or magnetism changes.

Where Raman data exist, \ac{dmc} agrees reasonably with other high-level theories and experiments. For graphene, we predict $\omega=73(1)$~cm$^{-1}$ compared to a previous DMC result of 83(7)~cm$^{-1}$ \cite{Mostaani2015} and experiment at 80(2)~cm$^{-1}$ \cite{Heinz2013}. For MoS$_2$ our \ac{dmc} value $\omega=36(4)$~cm$^{-1}$ agrees well with experiment at 40--41~cm$^{-1}$ \cite{Chen2015,Liangbo2017,Kim2021}.
WS$_2$ agrees within the \ac{dmc} uncertainty (31(5) vs.\ 33~cm$^{-1}$ \cite{Chen2015}), whereas MoSe$_2$, MoTe$_2$, and WSe$_2$ are several cm$^{-1}$ below the reported Raman values.
We did not find reports of rigid-layer breathing frequencies for h-BN, WTe$_2$, or the Cr trihalides to compare with our DMC results.
Ultralow-frequency Raman of few-layer h-BN maps the interlayer shear mode (52.5~cm$^{-1}$ in bulk; 46~cm$^{-1}$ at 3L) but not the breathing branch, which is Raman silent in the bulk\cite{Stenger2017}; the graphene 80~cm$^{-1}$ value was obtained from resonant overtones of the silent ZO$'$ mode \cite{Heinz2013}, a route closed for insulating h-BN.
Bulk $T_d$-WTe$_2$ shows an $A_1$ interlayer peak at 8--9~cm$^{-1}$ that overlaps a shear mode \cite{Ma2016,Kim2016WTe,Nema2023} and is not a rigid-layer analog of our $\omega=21(1)$~cm$^{-1}$.
Raman of CrI$_3$ and CrBr$_3$ reports intramolecular $A_g$/$E_g$ phonons from $\sim$50--280~cm$^{-1}$ \cite{DjurdjicMijin2018,Kozlenko2021} and the magnetism-coupled $\sim$129~cm$^{-1}$ $A_g$ Davydov splitting in few-layer CrI$_3$ \cite{Jin2020}, well above the rigid-layer well.

\subsection{Interlayer dipole moments}
We quantify interlayer-induced charge redistribution through the layer-resolved dipole moment
\begin{equation}
  \mu_z^{\mathrm{layer}} = \int_{z_{\min}}^{z_{\mathrm{vac}}}
  \rho(z)\,\bigl(z - z_{\mathrm{layer}}\bigr)\,\mathrm{d}z ,
\end{equation}
where $\rho(z)$ is the plane-averaged total bilayer density,
$z_{\mathrm{vac}}$ is the bilayer vacuum midpoint, and $z_{\mathrm{layer}}$ is an individual layer center.
The vacuum-bounded integral is evaluated on continuous sub-intervals
$[z_{\min}, z_{\mathrm{vac}})$ and $[z_{\mathrm{vac}}, z_{\max}]$ using linear interpolation of $\rho(z)$.
Bilayers are centered in the simulation cell along the $z$-axis, and periodic boundary contributions are neglected.
Figure~\ref{fig:dipole} summarizes $\lvert\mu_z^{\mathrm{layer}}\rvert$ as a function of nominal interlayer separation.

At representative short separations ($d=2.5$--3.0~\AA),
interlayer polarization is strongest in Cr-halides and weakest in graphene/\acs{hbn} stacks,
with \acp{tmd} occupying an intermediate regime.
Graphene and \acs{hbn} are grouped as single-sheet layers---each repeat unit is one atomic plane (graphene or monolayer \acs{hbn})---whereas \acs{tmd} and Cr-halide layers are multi-atom-thick.

For example, at $d=2.5$~\AA\ using PBE,
$\lvert\mu_z^{\mathrm{layer}}\rvert \approx 0.73~e$\,\AA\ (CrI$_3$),
$0.40~e$\,\AA\ (CrBr$_3$),
$0.24~e$\,\AA\ (MoTe$_2$),
$0.10~e$\,\AA\ (MoS$_2$),
and $\approx 0.07~e$\,\AA\ (\acs{hbn} and graphene).
CrI$_3$ exceeds CrBr$_3$ by nearly a factor of two at fixed $d$,
consistent with the larger polarizability and more diffuse halogen $p$ states of iodine.
Among \acp{tmd}, heavier chalcogens give larger $\lvert\mu_z^{\mathrm{layer}}\rvert$
(e.g.\ MoS$_2$ $<$ MoSe$_2$ $<$ MoTe$_2$ and WS$_2$ $<$ WSe$_2$ $<$ WTe$_2$ at $d=2.5$~\AA),
paralleling the halogen trend in CrX$_3$.
For CrX$_3$, all semilocal functionals and \acs{dmc} agree within a few percent at short range;
the material hierarchy is unchanged across these methods.
Magnetic stacking (AFM vs FM) changes $\lvert\mu_z^{\mathrm{layer}}\rvert$ by less than 1\% and does not affect the trends below \cite{supp}.

The $\lvert\mu_z^{\mathrm{layer}}\rvert$ decreases monotonically with increasing $d$ from 2.5 to 5~\AA\ for all materials,
reflecting the expected reduction of interlayer charge overlap as the layers decouple.
The systematic increase of interlayer polarization along
Br $\to$ I (CrX$_3$) and S $\to$ Se $\to$ Te (MX$_2$)
suggests that heavier, more polarizable anions enhance interlayer charge coupling at fixed $d$.
If such redistribution accompanies stronger interlayer exchange pathways,
it may contribute, together with other mechanisms, to the stronger interlayer magnetic coupling
observed in CrI$_3$ relative to CrBr$_3$.

For CrX$_3$ at short separations ($d \lesssim 3.5$~\AA), PBE and \acs{dmc} agree within a few percent
(e.g.\ CrI$_3$ at 2.5~\AA: $0.73$ vs.\ $0.77~e$\,\AA).
At larger $d$, where PBE predicts $\lvert\mu_z^{\mathrm{layer}}\rvert \to 0$ for all classes,
\acs{dmc} retains a long-ranged tail in both Cr-halides independent of the magnetic configuration:
at 5.0~\AA, $\lvert\mu_z^{\mathrm{layer}}\rvert \approx 0.007~e$\,\AA\ (PBE) vs.\ $0.082~e$\,\AA\ (\acs{dmc}) for CrBr$_3$,
and $0.014$ vs.\ $0.134~e$\,\AA\ for CrI$_3$,
whereas representative \acs{tmd} and single-sheet bilayers have $\lvert\mu_z^{\mathrm{layer}}\rvert \lesssim 0.015~e$\,\AA\ in both methods.
Thus, while both approaches predict rapid decay of interlayer coupling with separation,
many-body correlations in \acs{dmc} support weak but non-negligible interlayer charge redistribution throughout Cr-halide family. 
% \kscom{add Ganesh's comments on sliding multiferroicity} 

\section{Conclusion}
We have constructed a \ac{dmc} benchmark of interlayer binding for ten 2D bilayers spanning sp$^2$, \acs{tmd}, and magnetic chromium-trihalide chemistries with two magnetic configurations each.
From fitted $E_{\mathrm{b}}(d)$ curves we report equilibrium separations, binding energies, and breathing-mode frequencies, and we use them to assess widely used \ac{dft} approximations.
No single functional simultaneously matches \ac{dmc} for $d_0$, $E_{\mathrm{b}}$, and $\omega$ across the full set (Fig.~\ref{fig:dft_vs_dmc}); exact numerical agreement is not expected, but the pattern of failures is systematic.
As expected, functionals without any van der Waals correction perform poorly. Of the tests functionals that include these corrections,r$^2$SCAN--rVV10 gives the best overall agreement.
The tabulated bilayer curves and charge densities provide a transferable reference for developing and validating approximate interlayer methods.
% \kscom{add density related conclusion}

\section{Acknowledgments}
This work was supported by the \ac{doe}, Office of Science, Basic Energy Sciences, Materials Sciences and Engineering Division, as part of the Computational Materials Sciences Program and Center for Predictive Simulation of Functional Materials.
An award of computer time was provided by the \ac{incite} program. This research used resources of the \ac{alcf}, which is a \ac{doe} Office of Science User Facility supported under contract DE-AC02-06CH11357. This research also used resources of the \ac{olcf}, which is a \ac{doe} Office of Science User Facility supported under Contract DE-AC05-00OR22725.

\section*{Supporting Information}
Detailed numerical tables for equilibrium interlayer separation (Table~S1), binding energy (Table~S2), and out-of-plane breathing-mode frequency (Table~S3), the full set of binding-energy curves (Fig.~S1), and fit-quality diagnostics (Fig.~S2, Table~S4) are provided in the separate \acs{supinfo} document (\texttt{SI.pdf}).

\section*{Data Availability}
The tabulated data and binding curves in the \acs{supinfo}, together with ab initio electron densities available at DOI~\href{https://doi.org/10.13139/OLCF/2565002}{10.13139/OLCF/2565002}, provide a transferable reference for developing and validating approximate interlayer methods.

\bibliography{main}
\end{document}